\documentclass[aps,prl,10pt,a4paper,twocolumn,showpacs,noshowkeys,notitlepage]{revtex4-1}
\usepackage{amsmath}
\usepackage{amssymb}
\usepackage{braket}
\usepackage{tensor}
\usepackage{times}
\usepackage[dvipdfm]{graphicx}
\usepackage{hyperref}
\usepackage{dsfont}
\begin{document}
\title{A new view on spin reduced density matrices for relativistic particles}
\author{E. R. F. Taillebois}
\email{emiletaillebois@gmail.com}
\affiliation{Instituto de F\'{i}sica, Universidade Federal de Goi\'{a}s,%
 74.001-970, Goi\^{a}nia, Goi\'{a}s, Brazil}
\author{A. T. Avelar}
\email{ardiley@gmail.com}
\affiliation{Instituto de F\'{i}sica, Universidade Federal de Goi\'{a}s,%
 74.001-970, Goi\^{a}nia, Goi\'{a}s, Brazil}
\begin{abstract}
We present a new interpretation for reduced density matrices of secondary variables in relativistic systems via an analysis of Wigner's method to construct the irreducible unitary representations of the Poincar\'{e} group. We argue that the usual partial trace method used to obtain spin reduced matrices is not fully rigorous, however, employing our interpretation, similar effective reduced density matrices can be constructed. In addition, we show that our proposal is more useful than the usual one since we are not restricted only to the reduced density matrices that could be obtained by the ordinary partial trace method.
\end{abstract}
\pacs{03.67.-a,03.30.+p,03.65.Ta}
\maketitle


In the seminal work of Peres, Scudo, and Terno \cite{Peres2002}, the authors showed that the reduced density matrices (RDM) of spin obtained by the partial trace of the momentum degrees of freedom of massive particles do not have well defined transformation laws that connect their components in different inertial frames. This occurs due to the introduction of the relativistic symmetry in quantum mechanics that leads to the emergence of a hierarchy of dynamic variables. Relativistic transformation laws for primary variables (such as momentum) depend only on the Lorentz transformation acting on the system, while for secondary variables (such as spin and polarization) there is also a dependence on the momentum of the particle.

Since then, relativistic aspects of quantum information theory have attracted much attention mainly because of issues related to the behavior of entropy and quantum correlations for secondary variables in different frames. A large number of papers related to these subjects \cite{Gingrich2002,Peres2004,Kim2005,Landulfo2009,Czachor2003,Gingrich2003,Caban2005,Barlett2005,
Landulfo2010,Czachor1997,Czachor2010} deals with a regime where the number of particles does not change, free particles being described by irreducible unitary representations (IUR) of the universal cover of the Poincar\'{e} group $\tilde{\mathcal{P}}_{+}^{\uparrow}$ \cite{Wigner}. However, there are controversies about the validity and interpretation of the RDM obtained via the partial trace over the momentum degrees of freedom, as can be found in \cite{Czachor2005}, and more recently in \cite{Saldanha2012(1),Saldanha2012(2),Debarba2012}. In this letter we intend to solve these controversies.

We argue that, although the usual partial trace applied to secondary variables is not fully rigorous, effective RDM equivalent to those obtained by this method can be constructed if a correct interpretation is introduced. Furthermore, our approach also allows to construct different RDM that are able to describe results that the ordinary ones are incapable, as we will discuss later. Although the procedure and interpretation proposed here are valid for both massive and massless particles, for simplicity we will restrict ourselves to the case of massive particles. For the sake of clarity, in the following we will reproduce some steps from the literature \cite{Halpern,Tung,Weinberg} used to obtain the IUR of $\tilde{\mathcal{P}}_{+}^{\uparrow}$ that will be important for our argument. We will use $\hbar = 1$ and $c =1$.

Consider a 4-momentum $p$ and the homomorphism $L$ from $SL(2,\mathbb{C})$ to the restricted Lorentz group $\mathcal{L}_{+}^{\uparrow}$. The set
\begin{equation}
 \mathcal{O}(p) \equiv \{L(A)p|A\in SL(2,\mathbb{C})\}
\end{equation}
is called the orbit of $\mathcal{L}_{+}^{\uparrow}$ through $p$ and the 4-momenta in $\mathcal{O}(p)$ are said to be equivalent. Each IUR of $\tilde{\mathcal{P}}_{+}^{\uparrow}$ must have support concentrated in a single orbit and their structure can be determined by introducing an improper base $\{\ket{\mathbf{p},\alpha}\}$ of eigenvectors of the 4-momentum operator $P^{\mu}$ such that
\begin{equation}
 \braket{\mathbf{p},\alpha|\mathbf{q},\beta} = 2\omega_{\mathbf{q}}\delta(\mathbf{p}-\mathbf{q})\delta_{\alpha\beta}, \label{inn}
\end{equation}
where $\omega_{\mathbf{q}} = q^{0} = \sqrt{\|\mathbf{q}\|^{2} + m}$. The $\alpha$ labels in $\{\ket{\mathbf{p},\alpha}\}$ are the secondary variables and still have to be specified.

Given $q \in \mathcal{O}(p)$, the subgroup $G_{q} \subset SL(2,\mathbb{C})$ such that $L(M_q)q = q, \forall M_q \in G_q$, is called little group of $q$ . This definition implies that the infinite dimensional IUR of $\tilde{\mathcal{P}}_{+}^{\uparrow}$ are such that
\begin{equation}
 U(M_q)\ket{\mathbf{q},\alpha} = \sum_{\beta}Q^{q}_{\beta\alpha}(M_q)\ket{\mathbf{q},\beta}. \label{eq1}
\end{equation}
The $Q^{q}(M_q)$ matrices form a finite IUR of the little group $G_q$ and the $\alpha$ labels in $\ket{\mathbf{q},\alpha}$ are associated to base states of this representation in a finite dimensional Hilbert space $\mathcal{H}_q$. For this reason we will momentarily use the non-standard notation $\ket{\mathbf{q},\alpha_{q}}$. The little groups of equivalent 4-momenta are isomorphic and also are the Hilbert spaces $\mathcal{H}_q$.

To build the infinite dimensional IUR of $\tilde{\mathcal{P}}_{+}^{\uparrow}$ we need to introduce a rule to connect the $\alpha$ labels for any $q \in \mathcal{O}(p)$. Hence we choose a fundamental vector $k\in\mathcal{O}(p)$ and introduce a complementary set $\{C(p,k)\}$ such that for any $q\in\mathcal{O}(p)$ there is only one transformation $C(q,k)\in\{C(p,k)\}$ with $L(C(q,k))k = q$. For massive particles, $k$ is usually chosen to be the rest frame 4-momentum $(m,\mathbf{0})$ and the $\alpha$ labels are set to satisfy
\begin{equation}
 \mathbf{J}\ket{\mathbf{0},\alpha_{k}} = \sum_{\beta}(\mathbf{g})_{\beta\alpha}\ket{\mathbf{0},\beta_{k}},
\end{equation}
where $\mathbf{J} = (J_1,J_2,J_3)$ stand for the generators of the infinite dimensional IUR of $\tilde{\mathcal{P}}_{+}^{\uparrow}$ associated to the transformations in the subgroup $G_{k} = SU(2)$, and $\mathbf{g}$ the generators of the finite dimensional IUR of $G_{k}$ (as an example, for spin half particles, $\mathbf{g} = \vec{\sigma}/2$).

Once chosen a complementary set, we can define the base states for other 4-momenta equivalent to $k$ as
\begin{equation}
 \ket{\mathbf{p},\alpha_{p}}^{C} \equiv U[C(p,k)]\ket{\mathbf{0},\alpha_{k}}, \label{def}
\end{equation}
where $U[C(p,k)]$ represents the unitary operator associated to the transformation $C(p,k)$. The $C$ index is introduced due to the arbitrariness of the complementary set since any set $\{C^{\prime}(p,k)\}$ composed by transformations of the form $C^{\prime}(q,k) = C(q,k)M_{k}(q)$, with $M_{k}(q) \in G_{k}$ and $C(q,k)\in\{C(p,k)\}$, would serve as a complementary set. Therefore we can fully define the IUR of $\tilde{\mathcal{P}}_{+}^{\uparrow}$ by
\begin{equation}
 U(A)\ket{\mathbf{p},\alpha_{p}}^{C} = \sum_{\beta}Q_{\beta\alpha}^{k}(M_{k}^{C}(A,\mathbf{p}))\ket{L(A)\mathbf{p},\beta_{L(A)p}}^{C}, \nonumber
\end{equation}
where $A \in SL(2,\mathbb{C})$ and the generalized Wigner rotation reads
\begin{equation}
 M_{k}^{C}(A,\mathbf{p}) = C^{-1}(L(A)p,k)AC(p,k) . \label{wignerrot}
\end{equation}

This also allows to introduce the observables associated to the secondary variables. For each $\mathbf{p}$
\begin{equation}
 \boldsymbol{\mathcal{G}}_{C}(\mathbf{p})\ket{\mathbf{p},\alpha_{p}}^{C} = \sum_{\beta}(\mathbf{g})_{\beta\alpha}\ket{\mathbf{p},\beta_{p}}^{C}, \label{gcp}
\end{equation}
where $ \boldsymbol{\mathcal{G}}_{C}(\mathbf{p}) = U[C(p,k)]\mathbf{J}U^{\dagger}[C(p,k)]$ is a set of three generators for the infinite dimensional IUR of $\tilde{\mathcal{P}}_{+}^{\uparrow}$ associated to the transformations in the subgroup $G_p$. Although the $\alpha$ labels for different $q \in \mathcal{O}(p)$ are associated to different sets of operators, we can always construct an operator
\begin{equation}
 \boldsymbol{\mathcal{G}}_{C} = \sum_{\alpha}\int \frac{d\mathbf{p}}{2\omega_{\mathbf{p}}}\boldsymbol{\mathcal{G}}_{C}(\mathbf{p})\tensor*[]{\ket{\mathbf{p},\alpha_p}}{^C}\tensor*[^C]{\bra{\mathbf{p},\alpha_p}}{} \label{ger}
\end{equation}
such that
\begin{equation}
 \boldsymbol{\mathcal{G}}_{C}\ket{\mathbf{q},\alpha_{q}}^{C} = \sum_{\beta}(\mathbf{g})_{\beta\alpha}\ket{\mathbf{q},\beta_{q}}^{C}, \enskip \forall q \in \mathcal{O}(p).
 \label{gc}
\end{equation}
It is worth noting that the arbitrary dependence on the complementary set and the explicit dependence of the $\alpha$ labels on the momentum degrees of freedom in Eqs.(\ref{gcp}) to (\ref{gc}) make impossible to define any $ \boldsymbol{\mathcal{G}}_{C}$  as being $\mathds{1}\otimes\mathbf{g}$, as it was assumed explicitly or implicitly in several papers \cite{Peres2002,Gingrich2002,Peres2004,Kim2005,Landulfo2009}.

As an example we can choose the complementary set as being formed only by boosts, $C(p,k) \equiv B(p,k)$, leading to
\begin{equation}
 \boldsymbol{\mathcal{G}}_{B}(\mathbf{p}) \equiv \mathbf{S}(\mathbf{p}) = \frac{1}{m}\left(\mathbf{J}\omega_{\mathbf{p}} - \frac{(\mathbf{J}\cdot\mathbf{p})\mathbf{p}}{m+\omega_{\mathbf{p}}} - (\mathbf{p}\times\mathbf{K})\right),
\end{equation}
and the spin operators \cite{Terno2003}
\begin{equation}
 \boldsymbol{\mathcal{G}}_{B} \equiv \mathbf{S} = \frac{1}{m}\left(\mathbf{J}P^{0} - \frac{(\mathbf{J}\cdot\mathbf{P})\mathbf{P}}{m+P^{0}} - (\mathbf{P}\times\mathbf{K})\right),
\label{spin} 
\end{equation}
being $\mathbf{K}$ the boost generators. The base $\{\ket{\mathbf{p},\alpha_{p}}^{B}\}$ related to the complementary set $\{B(p,k)\}$ is called spin base.

Now we can look at the spin RDM obtained by the usual partial trace over the momentum degrees of freedom. A general density matrix for a one particle state can be written in the spin base as
\begin{equation}
 \rho = \sum_{\alpha,\beta}\int\int \frac{d\mathbf{p}}{2\omega_{\mathbf{p}}}\frac{d\mathbf{q}}{2\omega_{\mathbf{q}}}\rho_{\alpha\beta}^{B}(\mathbf{p},\mathbf{q})\tensor*[]{\ket{\mathbf{p},\alpha_p}}{^B}\tensor*[^B]{\bra{\mathbf{q},\beta_q}}{},
\end{equation}
where $\rho_{\alpha\beta}^{B}(\mathbf{p},\mathbf{q}) = \tensor*[^{B}]{\bra{\mathbf{p},\alpha_{p}}}{}\rho\tensor*[]{\ket{\mathbf{q},\beta_{q}}}{^{B}}$. Assuming that we can trace the momentum degrees of freedom, we obtain the spin RDM
\begin{equation}
 \rho_{spin} = \sum_{\alpha,\beta}\int\frac{d\mathbf{p}}{2\omega_{\mathbf{p}}} \rho_{\alpha\beta}^{B}(\mathbf{p},\mathbf{p})\tensor*[]{\ket{\alpha_p}}{^B}\tensor*[^B]{\bra{\beta_p}}{}. \label{red}
\end{equation}
If we compare two spin RDM obtained in this way, ``inner products'' of the form $\tensor*[^B]{\braket{\alpha_p|\beta_q}}{^B}$ will appear. These operations are not well-defined since the states belong to isomorphic but different Hilbert spaces. However, in works where the spin RDM obtained by the partial trace method are used, the momentum subindices are not explicit and the authors assume, explicitly or implicitly, that $\tensor*[^B]{\braket{\alpha_p|\beta_q}}{^B} = \delta_{\alpha\beta}$ \cite{Gingrich2002,Peres2004,Landulfo2009}. For this reason the usual partial trace method over the momentum degrees of freedom is not entirely stringent since we need to impose an 
assumption about $\tensor*[^B]{\braket{\alpha_p|\beta_q}}{^B}$, which is equivalent to the unnatural statement that there is a privileged complementary set. Observe that we are not contradicting (\ref{inn}) because $\tensor*[^{C}]{\braket{\alpha_{p}|\beta_{p}}}{^{C}} = \delta_{\alpha\beta}$ is 
well-defined since the states are in the same Hilbert space $\mathcal{H}_p$.

Two questions arise naturally. First, can we introduce a way to recover the results obtained by the partial trace method in a consistent way? Second, if we answer the first question in the affirmative, can the solution produce some kind of new result? To answer the first question we analyze the mean value of $\boldsymbol{\mathcal{G}}^C$ given by
\begin{equation}
 \mathrm{Tr}(\rho\boldsymbol{\mathcal{G}}_{C}) = \sum_{\alpha,\beta}\int \frac{d\mathbf{p}}{2\omega_{\mathbf{p}}}(\mathbf{g})_{\beta\alpha} \rho_{\alpha\beta}^{C}(\mathbf{p},\mathbf{p}) . \label{mean}
\end{equation}
We can introduce a Hilbert space $\mathcal{H}_C$ isomorphic to the $\mathcal{H}_p$'s and rewrite (\ref{mean}) as
\begin{equation}
 \mathrm{Tr}(\rho\boldsymbol{\mathcal{G}}_{C}) = \mathrm{Tr}_{\mathcal{H}_C}\left[\mathbf{g}\left( \sum_{\alpha\beta}\int \frac{d\mathbf{p}}{2\omega_{\mathbf{p}}} \rho_{\alpha\beta}^{C}(\mathbf{p},\mathbf{p}) \ket{\alpha}\bra{\beta} \right)\right], \nonumber
\end{equation}
where $\{\ket{\alpha}\}$ form a base in $\mathcal{H}_C$. This allows to define a density matrix
\begin{equation}
 \tau^{C} = \sum_{\alpha\beta}\ket{\alpha}\bra{\beta}\left(\int \frac{d\mathbf{p}}{2\omega_{\mathbf{p}}} \rho_{\alpha\beta}^{C}(\mathbf{p},\mathbf{p})\right) \label{tau}
\end{equation}
in the space of bounded linear operators $\mathcal{B}(\mathcal{H}_{C})$. If we choose $\{C(p,k)\} = \{B(p,k)\}$, the matrix $\tau^{B}$ will be equal to the spin RDM given by the customary partial trace method, but now the inner product $\braket{\alpha|\beta} = \delta_{\alpha\beta}$ will be well-defined since the states are in the same Hilbert space $\mathcal{H}_{B}$. Of course if we change the complementary set associated to $\boldsymbol{\mathcal{G}}_{C}$ the effective RDM furnished by our approach will be different and, in general, there will exist no well defined transformation law that connects these different matrices.

To finish answering the first question we have to give an interpretation to the effective RDM proposed. This is achieved observing that the mean value of any observable of the form $A_{C} = a_{0}I + \mathbf{a}\cdot\boldsymbol{\mathcal{G}}_{C}$, with $a_{0}$ and $\mathbf{a}$ real, can be written as
\begin{equation}
 \mathrm{Tr}(\rho A_{C}) = \mathrm{Tr}_{\mathcal{H}_C}(\tau^{C}a_{C}), \label{tr}
\end{equation}
where $a_{C} = a_{0}g_{0} + \mathbf{a}\cdot\mathbf{g}$ ($g_{0}$ is the identity operator). Therefore, for a given complementary set $\{C(p,k)\}$, the RDM obtained by our proposition give the statistical predictions for the results of measurements associated to observables that can be written as a linear combination of the components of $\boldsymbol{\mathcal{G}}_{C}$ and the identity operator $I$. Another way of introducing the effective density matrices and the interpretation presented above is to
define $\tau^{C}$ as the solution of (\ref{tr}), which leads to equation (\ref{tau}) due to the uniqueness property. This is the same approach used to introduce the partial trace as the only way to obtain consistent RDM for subsystems of ordinary composite systems \cite{Nielsen}. Therefore this approach allows to make explicit the inconsistency of the usual partial trace method in the context of secondary variables.

According to the presented interpretation the RDM that were obtained describe the statistical predictions of a system if we restrict the kind of measurements that we are able to perform over it. Furthermore, our approach permits to construct RDM for any choice of complementary set, as can be seen in (\ref{tau}), so we are not restricted only to the usual spin RDM. The only question that arise when we construct a RDM for an arbitrary complementary set is how to find an experimental setup that allows to measure the corresponding observables.

Next we will address the second question. To do it we analyze a model of spin measurement presented in \cite{Saldanha2012(1)} for which the results could not be described by the spin RDM obtained by the usual partial trace method. The proposed model consists of a neutral spin-1/2 particle that propagates with velocity
\begin{equation}
\mathbf{v} = v[\cos(\theta)\hat{\mathbf{x}}+\sin(\theta)\hat{\mathbf{y}}], \label{vel}
\end{equation}
and passes through two Stern-Gerlach (SG) apparatuses, the first (second) one in the $\hat{\mathbf{x}}$ ($\hat{\mathbf{y}}$) direction. The authors compute the expectation value of the measurement of the second apparatus after the first one has yielded an eigenvalue $+1/2$ for the spin component and show that if they consider the RDM obtained by the partial trace of the momentum degrees of freedom they cannot predict the results of such a model of spin measurement even when the particle is in a momentum eigenstate. Here we are going to show that the authors arrived at this conclusion only because the observable associated to their measurement is not a linear combination, independent of the momentum degrees of freedom, of the components of the spin operator given by (\ref{spin}) and, if our interpretation is used, correct effective RDM can be constructed in such a way that the results obtained by the authors are perfectly reproduced, circumventing in some sense the assumption made by them that ``the definition of a 
reduced density matrix for the particle spin is meaningless''. To this end we have to identify the observable that is being measured by their model of SG apparatus and associate it to a complementary set. 

Following the argument in \cite{Saldanha2012(1),Saldanha2012(2)} we consider that the quantization axis for the spin of a particle passing in the SG is given in the direction of the magnetic field in the rest frame of the particle by
\begin{equation}
 \hat{\mathbf{n}}(\hat{\mathbf{n}}_0,\mathbf{p}) = \frac{(\gamma + 1)\hat{\mathbf{n}}_0 -\gamma(\hat{\mathbf{n}}_0\cdot\mathbf{v})\mathbf{v}}{[1-(\hat{\mathbf{n}}_0\cdot\mathbf{v})^{2}]^{1/2}(\gamma+1)},
\end{equation}
where $\gamma = (1-v^{2})^{-\frac{1}{2}}$ and $\hat{\mathbf{n}}_0$ is the direction of the inhomogeneous magnetic field in the laboratory frame. Then we can introduce two rotations $R(\hat{\mathbf{n}}_0,\mathbf{p})$ and $R^{\prime}(\hat{\mathbf{n}}_0,\mathbf{p})$ such that
\begin{equation}
 \hat{\mathbf{n}}(\hat{\mathbf{n}}_0,\mathbf{p}) = R(\hat{\mathbf{n}}_0,\mathbf{p})\hat{\mathbf{n}}_0 = R^{\prime}(\hat{\mathbf{n}}_0,\mathbf{p})\hat{\mathbf{z}}.
 \label{rot}
\end{equation}
Note that there is some arbitrariness in the definition of these rotations since (\ref{rot}) does not uniquely define them, however this arbitrariness will not be important for our main results. In what follows we will abuse of the notation and use $R(\hat{\mathbf{n}}_{0},\mathbf{p})$ for both the $3\times3$ rotation matrices and the correspondent $SU(2)$ matrices.
Then the SG equipment will measure $\mathbf{J}\cdot \hat{\mathbf{n}}(\hat{\mathbf{n}}_0,\mathbf{p})$ in the rest frame of the particle, i.e. for a given momentum $\mathbf{p}$ the observable that will be measured is
\begin{align}
\begin{split}
 U[B(p,k)]\mathbf{J}\cdot \hat{\mathbf{n}}(\hat{\mathbf{n}}_0,\mathbf{p})U^{\dagger}[B(p,k)] = & \\  U[C(\hat{\mathbf{n}}_0;p,k)]\mathbf{J}\cdot\hat{\mathbf{n}}_0U^{\dagger}[C(\hat{\mathbf{n}}_0;p,k)] = & \\
 \mathbf{S}(\mathbf{p})\cdot\hat{\mathbf{n}}(\hat{\mathbf{n}}_0,\mathbf{p}) & \label{op}
\end{split}
\end{align}
where $C(\hat{\mathbf{n}}_0;p,k)= B(p,k)R(\hat{\mathbf{n}}_0,\mathbf{p})$.

Henceforth the momentum subindices for base states will be omitted. We define $\ket{\mathbf{p},\alpha}^{C_{n_0}}$ as the eigenstate of the operator (\ref{op}) such that
\begin{equation}
 \mathbf{S}(\mathbf{p})\cdot\hat{\mathbf{n}}(\hat{\mathbf{n}}_0,\mathbf{p})\ket{\mathbf{p},\alpha}^{C_{n_0}} = \alpha\ket{\mathbf{p},\alpha}^{C_{n_0}}.
\end{equation}
Thus
\begin{equation}
 \ket{\mathbf{p},\alpha}^{C_{n_0}} = U[C(\hat{\mathbf{n}}_0;p,k)]\ket{\mathbf{0},\alpha}^{C_{n_0}},
\end{equation}
where $\ket{\mathbf{0},\alpha}^{C_{n_0}}$ is such that $ \mathbf{J}\cdot\hat{\mathbf{n}}_0 \ket{\mathbf{0},\alpha}^{C} = \alpha \ket{\mathbf{0},\alpha}^{C}$.

If we want to describe the observable associated to the experiment for any momentum we use (\ref{ger}) to write the operator
\begin{equation}
 \mathcal{G}^{3}_{C_{n_0}} = \sum_{\alpha}\int \frac{d\mathbf{p}}{2\omega_{\mathbf{p}}}\mathbf{S}(\mathbf{p})\cdot\hat{\mathbf{n}}(\hat{\mathbf{n}}_0,\mathbf{p})\tensor*[]{\ket{\mathbf{p},\alpha}}{^{C_{n_0}}}\tensor*[^{C_{n_0}}]{\bra{\mathbf{p},\alpha}}{}, \nonumber
\end{equation}
which represents the observable measured by the SG apparatus and is clearly not a linear combination of the components of the spin operator in (\ref{spin}).

Now we suppose an eigenstate of momentum with eigenvalue $+1/2$ for a SG-$\hat{\mathbf{x}}$ measure, i.e. $ \ket{\psi} = \ket{\mathbf{p},+1/2}^{C_x}$, where $C_x$ is related to the complementary set composed by operators of the form
$C(\hat{\mathbf{x}};p,k) = B(p,k)R(\hat{\mathbf{x}},\mathbf{p})$, and $\mathbf{p} = \gamma m \mathbf{v}$, with $\mathbf{v}$ given by (\ref{vel}).
We want to evaluate the expectation value of a measure realized by an SG-$\hat{\mathbf{y}}$ apparatus over this state. To do so we will need the following relations
\begin{align}
 \ket{\mathbf{p},\alpha}^{C_x} & = \sum_{\beta}Q^{k}_{\beta\alpha}[R^{\prime}(\hat{\mathbf{x}},\mathbf{p})]\ket{\mathbf{p},\beta}^{B}, \\
 \ket{\mathbf{p},\alpha}^{C_y} & = \sum_{\beta}Q^{k}_{\beta\alpha}[R^{\prime}(\hat{\mathbf{y}},\mathbf{p})]\ket{\mathbf{p},\beta}^{B},
\end{align}
where $C_y$ is associated to the complementary set of the second SG, formed by transformations of the form
\begin{equation}
 C(\hat{\mathbf{y}};p,k) = B(p,k)R(\hat{\mathbf{y}},\mathbf{p}).
\end{equation}

For the momenta we are considering the rotation matrices $R^{\prime}(\hat{\mathbf{x}},\mathbf{p})$ and $R^{\prime}(\hat{\mathbf{y}},\mathbf{p})$ are
\begin{equation}
  R^{\prime}(i,\mathbf{p}) = \frac{\sqrt{2}}{2}
 \begin{pmatrix}
  e^{i(\chi_{i}+\phi_{i})/2} & i e^{i(\chi_{i}-\phi_{i})/2} \\
  i e^{-i(\chi_{i}-\phi_{i})/2} & e^{-i(\chi_{i}+\phi_{i})/2}
 \end{pmatrix}, \label{matrot}
\end{equation}
with $i = \hat{\mathbf{x}}, \hat{\mathbf{y}}$ and
\begin{align}
  \begin{split}
 \cos(\chi_i) & = \hat{\mathbf{n}}(i,\mathbf{p})\cdot\hat{\mathbf{y}} \\
 \sin(\chi_i) & = \hat{\mathbf{n}}(i,\mathbf{p})\cdot\hat{\mathbf{x}}
  \end{split}.
 \end{align}
The angle $\phi_i$ is associated to the degree of arbitrariness that was mentioned earlier.
For spin half particles $Q^{k}(R^{\prime}(\hat{\mathbf{n}}_{0},\mathbf{p})) = R^{\prime}(\hat{\mathbf{n}}_{0},\mathbf{p})$
and, using the approach presented here, we can construct the density matrix that describes the result of the measurements of the SG-$\hat{\mathbf{y}}$ apparatus:
\begin{equation}
  \begin{footnotesize}
  \tau^{C_y} = \frac{1}{2}
 \begin{pmatrix}
  1 + \hat{\mathbf{n}}(\hat{\mathbf{x}},\mathbf{p})\cdot\hat{\mathbf{n}}(\hat{\mathbf{y}},\mathbf{p}) & e^{-i\phi_{y}}[\hat{\mathbf{n}}(\hat{\mathbf{x}},\mathbf{p})\times\hat{\mathbf{n}}(\hat{\mathbf{y}},\mathbf{p})]\cdot\hat{\mathbf{z}} \\
  e^{i\phi_{y}}[\hat{\mathbf{n}}(\hat{\mathbf{x}},\mathbf{p})\times\hat{\mathbf{n}}(\hat{\mathbf{y}},\mathbf{p})]\cdot\hat{\mathbf{z}} & 1 - \hat{\mathbf{n}}(\hat{\mathbf{x}},\mathbf{p})\cdot\hat{\mathbf{n}}(\hat{\mathbf{y}},\mathbf{p})
 \end{pmatrix}, \label{taufinal} \nonumber
  \end{footnotesize}
\end{equation}
It is important to note that the momentum only appears explicitly in $\tau^{C_{y}}$ because we are dealing with a momentum eigenstate. Otherwise, integrals over the momentum degrees of freedom would appear inside the matrix. We also emphasize that the dependence on $\hat{\mathbf{n}}(i,\mathbf{p})$, with $i=\hat{\mathbf{x}},\hat{\mathbf{y}}$, is due to the effective feature of the RDM.

The observable associated to the measurement of this SG is $\mathcal{G}^{C_y}_{3}$ and, therefore, for the correspondent effective reduced density matrix, this observable will be described by $g_3 = \sigma_3/2$, where $\sigma_3$ is the usual Pauli matrix. Then the expectation value for the given measurement is
\begin{equation}
 \mathrm{Tr}_{\mathcal{H}_{C_y}}\left(\frac{\sigma_3}{2}\tau^{C_y}\right)  =\frac{-v^{2}\cos\theta\sin\theta}{2\sqrt{(1-v^{2}\cos^{2}\theta)(1-v^{2}\sin^{2}\theta)}}, \nonumber
\end{equation}
in accordance with \cite{Saldanha2012(1)}.

Finally, we need to analyze the importance of the present formalism for relativistic quantum information theory. To achieve this goal we have to answer a last question: given an apparatus described by observables $\boldsymbol{\mathcal{G}}_{C}$ and RDM associated to the corresponding complementary set, what kind of Lorentz transformation can we perform over the apparatus and still use the same RDM to describe the system? Assuming that we apply a transformation $U(A)$ over the apparatus and reminding that
\begin{equation}
\boldsymbol{\mathcal{G}}_{C} = (\mathcal{G}_{C}^{1},\mathcal{G}_{C}^{2},\mathcal{G}_{C}^{3}) = (\mathcal{G}_{C}^{23},\mathcal{G}_{C}^{31},\mathcal{G}_{C}^{12}),
\end{equation}
the new observable associated to the measurement will be given by:
\begin{widetext}
\begin{equation}
   U(A)\mathcal{G}^{l}_{C}U^{\dagger}(A) = U(A)\mathcal{G}^{mn}_{C}U^{\dagger}(A)   
  = \sum_{\alpha}\int \frac{d\mathbf{p}}{2\omega_{\mathbf{p}}}\tensor[]{M_{k}^{C}(A^{-1},\mathbf{p})}{^{m}_{i}}  \tensor[] {M_{k}^{C}(A^{-1},\mathbf{p})}{^{n}_{j}}  \mathcal{G}^{ij}_{C}(\mathbf{p})\tensor[]{\ket{\mathbf{p},\alpha}}{^C}\tensor[^C]{\bra{\mathbf{p},\alpha}}{},
\end{equation}
\end{widetext}
the sum over the latin indexes $i,j$ being from 1 to 3. This new observable will be a linear combination of the components of $\boldsymbol{\mathcal{G}}_{C}$ only if $M_{k}^{C}(A^{-1},\mathbf{p}) = M_{k}^{C}(A^{-1})$, in which case
\begin{equation}
   U(A)\mathcal{G}^{jk}_{C}U^{\dagger}(A) =
   \tensor[]{M_{k}^{C}(A^{-1})}{^{m}_{i}}\tensor[]{M_{k}^{C}(A^{-1})}{^{n}_{j}}\mathcal{G}^{ij}_{C}.\label{last}
\end{equation}
For quantum information theory the RDM formalism will be particularly interesting when the set of transformations that satisfy (\ref{last}) for a given complementary set is the group of rotations, since it will be possible to rotate the equipment and still use the same RDM to describe the measurements. This will not be the case for every complementary set, as we can see from the  previous example. An example of  complementary set for which the rotation group satisfies (\ref{last}) is the one composed only by boost (spin base).

To summarize in this letter we have addressed the issue of the adequacy of the partial trace for the construction of RDM of secondary variables, in particular spin. We showed that, despite being constructed by a not quite stringent method, the  RDM obtained by the partial trace over the momentum degrees of freedom can be recovered by introducing an adequate interpretation and abandoning the usual partial trace. This new interpretation shows that there is no unique definition for the RDM of spin: depending on the set of observables associated to the measure apparatus, the associated RDM will be different. Thus we have presented a way to unify without inconsistencies the usual results in literature \cite{Peres2002,Gingrich2002,Gingrich2003,Peres2004,Barlett2005,Kim2005,Landulfo2009,Landulfo2010} with the recent criticisms in \cite{Saldanha2012(1),Saldanha2012(2)}. It is worth noting the importance of our result for quantum information theory since, once a good spin measurement will be defined, it will be 
important to know how to construct RDM associated to this measurement and if these RDM are going to be invariant by rotation of the equipment. As a final remark we stress that the present formalism may be useful to solve other issues related to relavistic quantum information, such as the spin tomography of relativistic massive particles, a problem that was recently raised in \cite{Palmer}. 

We are grateful to Daniel Terno, Pablo Saldanha, and Marek Czachor for enlightening discussions. We thank the financial support provided by Brazilian agencies CNPq, CAPES and the Instituto Nacional de Ci\^encia e Tecnologia - Informa\c c\~ao Qu\^ antica (INCT-IQ). 
\bibliography{TIQR}
\end{document}